%% file: tristab.tex
\documentclass[11pt,twoside,letterpaper]{article}                                                                                                              

\usepackage{latexsym}
\usepackage{amsmath}
\usepackage{dcolumn}
\usepackage{enumitem}
\usepackage{authblk}

\usepackage[margin=1in]{geometry}

\usepackage{graphicx}
\usepackage{amssymb}
\usepackage{bm}
\usepackage{color}
\usepackage{soul}
\usepackage{hyperref}
\usepackage{enumerate}
\usepackage{stackengine}
\usepackage{algorithm}
\usepackage[noend]{algpseudocode}
\usepackage[figuresright]{rotating}
\usepackage{orcidlink}

\definecolor{green1}{rgb}{0.0, 0.5, 0.0}

\input kigou.tex

\newcommand{\GA}{\alpha}
\newcommand{\GB}{\beta}
\newcommand{\GG}{\gamma}
\newcommand{\GD}{\delta}

\newcommand{\GR}{\rho}

\newcommand{\GC}{\psi}

\newcommand{\GZ}{\zeta}
\newcommand{\GP}{\phi}

\newcommand{\GU}{\theta}

\newcommand{\pd}{\partial}

\newcommand{\be}{\begin{equation}}
\newcommand{\ee}{\end{equation}}

\begin{document}

\title{General Relativistic Stability and Gravitational Wave Content of Rotating Triaxial Neutron Stars}

\author[1]{Yufeng Luo}
\author[2,3,4]{Antonios Tsokaros}
\author[3,2]{Roland Haas}
\author[5]{K{\= o}ji Ury{\= u}}

\affil[1]{Department of Physics and Astronomy, University of Wyoming, Laramie, Wyoming, 82071, USA}
\affil[2]{Department of Physics, University of Illinois at Urbana-Champaign, Urbana, IL 61801, USA}
\affil[3]{National Center for Supercomputing Applications, University of Illinois at Urbana-Champaign, Urbana, IL 61801, USA}
\affil[4]{Research Center for Astronomy and Applied Mathematics, Academy of Athens, Athens 11527, Greece}
\affil[5]{University of the Ryukyus, Department of Physics, Senbaru, Nishihara, Okinawa 903-0213, Japan}

\maketitle

\begin{abstract}
Triaxial neutron stars can be sources of 
continuous gravitational radiation detectable by ground-based interferometers. The amplitude of the
emitted gravitational wave can be greatly affected by the state of the hydrodynamical fluid flow inside the neutron star.
In this work we examine the most triaxial models along two sequences of constant 
rest mass, confirming their dynamical stability. We also study the
response of a triaxial figure of quasiequilibrium under a variety of perturbations
that lead to different fluid flows. Starting from the general relativistic compressible analog of the 
Newtonian Jacobi ellipsoid, we perform simulations of Dedekind-type flows. We find that in some cases
the triaxial neutron star resembles a Riemann-S-type ellipsoid with minor rotation and gravitational
wave emission as it evolves towards axisymmetry. The present results highlight the importance of
understanding the fluid flow in the interior of a neutron star in terms of its gravitational wave content.
\end{abstract}

\section{Introduction}

One of the most profound predictions of general relativity is that a system
which possesses time varying multipole moments higher than a quadrupole,
generates gravitational waves.
The most common systems that satisfy such criterion are the ones that not
symmetric about their rotation axis, with prime examples being those of binary
compact objects.  Therefore it is not a surprise that the first direct
detection of a gravitational wave came from a binary black hole
\cite{Abbott2016a,Abbott2016fw}. In the first three observational periods
(O1-O3), the LIGO/Virgo \cite{Aasi:2014,Acernese:2014} collaboration has
discovered gravitational waves from almost 100 binaries
\cite{LIGO_GWTC1,LIGO_GWTC2,LIGO_GWTC3} including two binary neutron stars
\cite{GW170817prl,Abbott2017b,Abbott2018b,LIGOScientific_GW190425} and two
black hole-neutron star mergers \cite{LIGOScientific:2021qlt}.  Another
exciting possibility is to detect gravitational waves from a single neutron
star that exhibits some kind of asymmetry
\cite{Haskell:2014nna,Riles2022,Pagliaro:2023bvi,Wette2023}.  Although such
gravitational waves are much weaker than the ones emerging from a binary system
(and this is one of the reasons that they have not been detected yet), they have the
potential of providing important information regarding the nature of a neutron
star such as regarding various fluid instabilities or its elastic, thermal and magnetic
characteristics.

A hydrodynamical instability is one well-known mechanism that can produce
nonaxisymmetric neutron stars which emit gravitational waves \cite{Friedman2012}. 
One important parameter that characterizes unstable rotating
neutron stars is $\beta :=T/|W|$, where $T$ is the rotational kinetic energy
and $|W|$ the gravitational binding energy \cite{Friedman2012,Baumgarte2010}.
As the rotation of the star increases, there are two critical points 
(nonaxisymmetric instabilities) that are
associated with two different physical mechanisms. In the presence of some
dissipative mechanism such as viscosity or gravitational radiation, at
$\beta=\beta_s$ the star becomes secularly unstable to a bar-mode deformation.
The timescale of this instability is set by the dissipation and is much longer
than the dynamical (free-fall) timescale. At even higher rotation rates, when
$\beta=\beta_d > \beta_s$ the star becomes dynamically unstable to a bar-mode
deformation.  This instability emerges regarless of any possible dissipation
and its growth is set by the dynamical timescale. For incompressible stars in
Newtonian gravity $\GB_s^{\rm Newt}=0.1375$ and $\GB_d^{\rm Newt}=0.2738$ \cite{Chandrasekhar69c}.
Although the values of $\GB$ at these critical points can change in general
relativity, with compressible equations of state and differential rotation, the
overall idea (existence of distinct secular and dynamical instability points)
remains\footnote{Nonaxisymmetric instabilities for values of $\GB$ as low as
0.01 have also been found \cite{Shibata:2002mr,Shibata:2003yj}. 
These so-called shear instabilities depend
on $\GB$ and the degree of differential rotation \cite{Watts2005}.}.

The broadbrush picture above can be further refined by the fact that there are two
categories of secular instabilities: i) The viscosity-driven instability which
as the name suggests manifests itself in the presence of viscous dissipation
\cite{Roberts1963}, and ii) the Chandrasekhar-Friedman-Schutz (CFS) instability
which is driven by gravitational radiation reaction
\cite{Chandrasekhar:1970,Friedman78,Friedman1978c}.  For Newtonian
incompressible fluids, an axisymmetric rotating body is described by a
Maclaurin spheroid \cite{Chandrasekhar69c}, an oblate spheroid having
$R_x=R_y\neq R_z$.  At the point of secular instability, when $\GB=0.1375$, two
families of triaxial ($R_x\neq R_y\neq R_z$) solutions emarge: a) The Jacobi
ellipsoids, which are uniform rotating ellipsoidal figures of equilibrium in
the inertial frame, and thus emit gravitational waves, and b) the Dedekind
ellipsoids, which are ellipsoidal figures of equilibrium stationary in the
inertial frame, and therefore do not emit gravitational waves\footnote{That
does not mean that the evolution along the Dedekind sequence does not produce
gravitational waves.}. The Dedekind
ellipsoids have constant vorticity and nonzero internal fluid circulation.
Equilibrium solutions (a) and (b) are related to the processes (i) and (ii)
respectively as follows \cite{Lai93,Lai95,Christodoulou95i}. Viscosity dissipates mechanical
energy but conserves angular momentum and a Jacobi ellipsoid has less
mechanical energy, $T+W$, than a Maclaurin spheroid of the same rest mass and
angular momentum. Thus the viscous-driven evolution (i) of an unstable
Maclaurin spheroid would proceed towards a Jacobi ellipsoid (a). On the other
hand, gravitational radiation preserves circulation along any closed path on a
plane parallel to the equator, but not angular momentum. A Dedekind ellipsoid
has less mechanical energy, than a Maclaurin spheroid of the same rest mass and
circulation. Thus, in the absence of viscosity, the CFS-driven (ii) evolution
of an unstable Maclaurin spheroid would proceed towards a Dedekind ellipsoid
(b).  The presence of both viscosity and gravitational radiation tends to
stabilize the star against these competing mechanisms \cite{Lindblom1977}.  In
the limit where the gravitational wave timescale equals the viscous timescale,
the Maclaurin spheroid is secularly stable all the way to the dynamical
instability point.

One important difference between the viscosity-driven instability and the CFS
instability is that the latter is generic while the former is absent in sufficiently
slowly rotating stars. In addition the  viscosity-driven instability
emerges only for sufficiently stiff equations of state in which the bifurcation
point exists before the mass-shedding (Keplerian) limit. In Newtonian gravity 
with a polytropic equation of state, $p=k\rho^\Gamma$, the triaxial sequence
exists only if $\Gamma \gtrapprox 2.24$ \cite{James64}. In general relativity 
the critical adiabatic index does not change significantly but slightly increases
$\sim 2.4$ \cite{Ipser1981,Bonazzola1996b,Skinner1996,Bonazzola1998c}. 
At the same time the critical parameter $\GB_s$ also increases relative to the
Newtonian value ($0.1375$) by a factor that depends on the compactness of the
neutron star \cite{Rosinska2002a}. On the other hand, the CFS instability
becomes stronger in general relativity and sets in at $\GB<0.1375$
\cite{Stergioulas98a,Morsink99} so that the two instabilities no longer occur at 
the same value of $\GB_s$. 

Sequences of triaxial solutions in general relativity have been
investigated in \cite{Huang08,Uryu2016a,Uryu2016b} using a select set of stiff
equations of state.  It was found that the triaxial sequence becomes shorter (a
smaller deformation is allowed) as the compactness increases, while
supramassive \cite{Cook92b} triaxial equilibria are possible, depending on the
equation of state. In \cite{tsokaros2017} the first full general relativistic
simulations of triaxial uniformly rotating neutron stars have been performed,
and the dynamical stability of certain normal and supramassive solutions was
established.  It was found that all triaxial models evolve toward axisymmetry,
maintaining their uniform rotation, while losing their triaxiality through gravitational wave
emission.  Similar results were reported in \cite{Zhou2017xhf} where triaxial
quark stars (having finite surface density) were evolved in general
relativity. 

In this work we investigate the fate and stability of triaxial models against a
variety of perturbations. First we establish the dynamical stability of the most
triaxial figure of quasiequilibrium along two constant rest-mass sequences,
one that corresponds to compactness $0.1$, and another one that corresponds to
compactness $0.19$. Second, by replacing the Jacobi-like velocity flow with a
Dedekind-like one we explore the fate of the resulting ellipsoidal neutron star.
We find that in some cases this procedure leads to a Riemann-S-type ellipsoidal
figure of quasiequilibrium that barely rotates while largely preserving its
nonaxisymmetric shape. This object emits gravitational waves whose amplitude is
approximately $20\%$ the one coming from the original triaxial neutron star
as it evolves towards axisymmetry, and highlights the importance of the fluid flow 
in accurate gravitational wave analysis.


Here we employ geometric units in which $G=c=M_\odot=1$, unless stated
otherwise. Greek indices denote spacetime dimensions (0,1,2,3), while
Latin indices denote spatial ones (1,2,3).

\section{Numerical methods and model parameters}
\label{sec:nm}

For the construction of the initial models we use the \cocal{} code
as described in \cite{Huang08,Uryu2016a,Uryu2016b} while for the 
evolution we use the \etoolkit{} \cite{EinsteinToolkit:2020_05,Etienne2015,Noble2006,Schnetter:2003rb,Dreyer:2002mx}.
Below we summarize the most
import features of our initial data models and their evolutions.

\subsection{Initial data}
\label{sec:id}

We construct uniformly rotating triaxial neutron stars having angular velocity $\Omega$ and velocity with respect to the inertial frame
$v^i = \Omega\phi^i=\Omega (-y,x,0)$. The fluid's 4-velocity can be written as 
\begin{equation}
u^\alpha = u^t k^\alpha = u^t (t^\alpha + v^\alpha) \ ,
\label{eq:4vel}
\end{equation} 
where $u^t$ is a scalar. The spacetime of the rotating star possesses a helical Killing vector, $k^\alpha$, where
\begin{equation}
k^\alpha = t^\alpha + \Omega \phi^\alpha,
\label{eq:hkv}
\end{equation}
with the fluid variables being Lie-dragged along $k^\alpha$,
\begin{equation}
\mathcal{L}_{\bf k}(hu_\alpha) = \mathcal{L}_{\bf k}\rho = \mathcal{L}_{\bf k} s = 0.
\label{eq:sym}
\end{equation}
Here $\rho,\ h,\ s$ are the rest-mass density, enthalpy, and the
entropy per unit rest-mass. We have $\rho h=\epsilon+p$, where $\epsilon$ is the total energy density
and $p$ is the pressure. 

In order to ensure the existence of triaxial uniformly rotating models we use a stiff polytropic equation of state
with $\Gamma=4$. For the polytropic constant we chose $k=1$ in $G=c=M_\odot=1$. Similar to \cite{Uryu2016b,tsokaros2017}
the value of $\Gamma$ used is simply to prove a point of principle, rather than to address physical EOS parameters.

The models have been computed with the \cocal{} code, a second-order
finite-difference code whose methods are explained, for example, in
\cite{Uryu2016a,Tsokaros2022r}. For computational convenience, 
we employ the Isenberg-Wilson-Mathews (IWM) formulation \cite{Isenberg08,Wilson89,Wilson95,Wilson96,Tsokaros2022r}.
Therefore the 3-metric is $\GG_{ij}=\GC^4 f_{ij}$, where $\GC$ is the conformal factor and
$f_{ij}$ the flat metric. The unknown gravitational variables in the 3+1 formulation are
the lapse $\GA$, the shift $\GB^i$, and the conformal factor $\GC$. In the \cocal{} code
the full system of equations (wavless formulation) is also used but
the differences from the conformally flat IWM scheme are small \cite{Uryu2016a}.
A number of diagnostics are used to describe the initial solutions and
explicit formulae are given in the appendix of \cite{Uryu2016a}.  
The most important diagnostics are:
1) The angular momentum of the triaxial star $J$ (where $J$ is the
  Arnowitt-Deser-Misner (ADM) angular momentum), which is computed via a
surface integral at infinity or a volume integral over the spacelike
hypersurface. 
2) The kinetic energy, which is defined as $T:=\frac{1}{2}J\Omega$
(although we are not in axisymmetry we still use this formula because it is gauge-invariant), and 
3) the gravitational potential energy, which is
defined as $W:=M - M_p - T$. Here $M$ is the
(ADM) mass and $M_p$ is the proper mass (rest-mass plus internal energy) of
the star (see e.g. \cite{Friedman1986}).  These expressions are used then to
compute the rotation parameter $\beta=T/|W|$. 
We also define the moment of inertial as $I:=J/\Omega$. 

To measure of accuracy of the initial data we use two diagnostics: The first one is the difference between
the Komar and ADM mass \cite{Uryu2016a},
\be 
\GD M = \frac{|M_{\rm K}-M|}{M_{\rm K}} \,.
\label{eq:KmADM}
\ee
For stationary and asymptotically flat spacetimes $M_{\rm K}=M$ \cite{Beig1978}.
The second diagnostic is the relativistic virial equation \cite{Gourgoulhon1994}.
In our calculations both diagnostics are $ \sim O(10^{-4})$.

\begin{figure}[H]
\begin{center}
\includegraphics[width=6.5 cm]{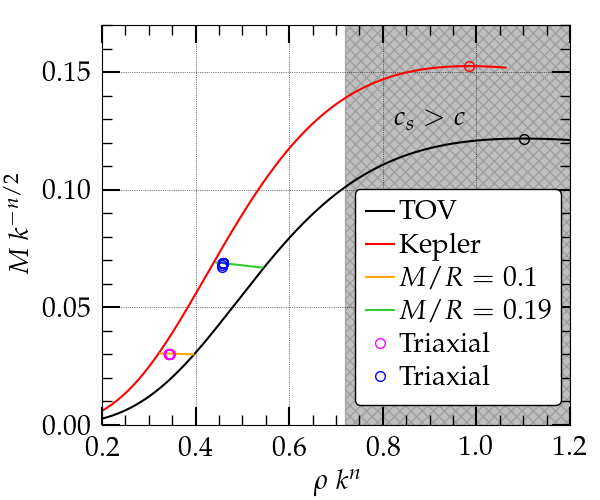}
\includegraphics[width=6.5 cm]{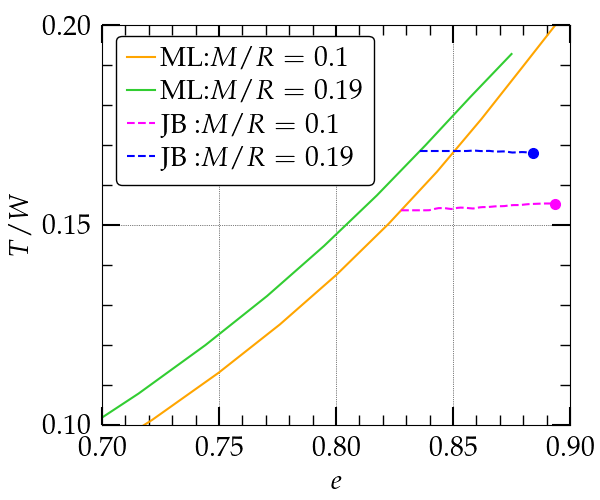}
\caption{Left panel: Mass versus rest-mass density for the spherical (black line) and mass-shedding (red line) limits. Also
plotted are sequences of constant rest-mass (green and orange lines) for compactions $(M/R)_s=0.19$ and $0.1$. Magenta and blue circles denote
triaxial models. Right panel: $T/W$ versus the eccentricity $e$ for sequences of Mclaurin (ML-orange, 
ML-green) and Jacobi (JB-blue, JB-magenta) type ellipsoids for compactions $(M/R)_s=0.1,\ 0.19$. 
Solid magenta and blue circles are models C010s17 and C019s08 respectively.}
\label{fig:tov}
\end{center}
\end{figure}

We start our calculations by computing two sequences of axisymmetric rotating neutron stars having constant rest mass that correspond to spherical compactions $(M/R)_s=0.1$ and $0.19$. 
These sequences which are shown with orange ($(M/R)_s=0.1$) 
and green ($(M/R)_s=0.19$) color in Fig. \ref{fig:tov}, are the analogues of the Maclaurin spheroids
in Neutonian gravity \cite{Chandrasekhar69c}. In the left panel of Fig. \ref{fig:tov}, the mass is plotted 
against the central density, while in the right panel we plot the the rotation parameter $\beta$ versus the 
eccentricity $e:=\sqrt{1-(\bar{R}_z/\bar{R}_x)^2}$ with respect to the proper radii. Note that the mass and
the density can by rescaled to any number using the polytropic constant $k$\footnote{In geometric units,
$k^{n/2}$, where $\Gamma=1+1/n$ and $n$ the polytropic index, has units of length.}, 
hence the axes in the left panel are normalized accordingly.
In the left panel of Fig. \ref{fig:tov} we also show  the sequence of spherically symmetric Tolman-Oppenheimer-Volkov (TOV) 
solutions (black curve), and the sequence of maximally uniformly rotating (at the mass-shedding limit, also called the Kepler limit) 
solutions (red curve). With black (red) circle we denote the solution of maximum nonrotating (uniformly rotating) mass.
The shaded area corresponds to densities where the speed of sound $c_s$ is larger than the speed of light. All the solutions
used in this work are causal.

For sufficiently high rotation rate ($\GB$) a second branch of solutions appears. These are triaxial solutions ($R_x\neq R_y\neq R_z$)
that correspond to the Newtonian Jacobi ellipsoids \cite{Chandrasekhar69c}. The two sequences that correspond to $(M/R)_s=0.1$ and $0.19$
are shown with blue ($(M/R)_s=0.1$) and magenta ($(M/R)_s=0.19$) colors. In the left panel of Fig. \ref{fig:tov} all triaxial solutions that
correspond to each sequence have masses and densities very close to each other so they appear as a single triangle point (magenta or blue). 
In the right panel though, the triaxial sequences are clearly seen. For a fixed eccentricity a triaxial model has less $T/|W|$ than
the corresponding axisymmetric model. In particular for a fixed eccentricity the triaxial solution has less gravitational (ADM) mass $M$
(thus more negative gravitational potential energy $W$),
angular momentum $J$ , angular velocity $\Omega$, and moment of inertia $I$ than the corresponding axisymmetric model. 
Therefore it has less kinetic energy too. On the other hand it has larger proper mass, hence less total energy $T+W=M-M_p$\footnote{$T+W$ 
is more negative for the triaxial solution.} and thus it is the preferred figure of equilibrium.

\begin{table*}
\begin{tabular}{cccccccccccccc}
\hline
\hline
Model & $\rho$ & $R_x$ & $R_z/R_x$ & $R_y/R_x$ & $e$ & $\Omega M$   \\
\hline
C010s17 & $0.3430$ & $0.4421$ & $0.4444$ & $0.6875$ & $0.8918$ & $0.01808$    \\  
C019s08 & $0.4583$ & $0.4476$ & $0.4619$ & $0.7813$ & $0.8778$ & $0.05098$  \\  
\hline
\hline
Model & $M$ & $M_0$ & $J/M^2$ & $(M/R)_s$ & $T/|W|$ & $I(\times 10^{-3})$ \\
\hline
C010s17 & $0.03042$ & $0.03193$ & $1.117$  & $0.1$  & $0.1547$ & $1.740$   \\  
C019s08 & $0.06888$ & $0.07578$ & $0.9011$ & $0.19$ & $0.1676$ & $5.781$  \\  
\hline
\hline
\end{tabular}
\caption{Initial data models C010s17 and C019s08 used in this  work.
Here $\rho$, $R_i$, $e=\sqrt{1-(\bar{R}_z/\bar{R}_x)^2}$,
$\Omega$, $M$, $M_0$, $(M/R)_s$, $T/|W|$, $I$, are the
central rest-mass density, the coordinate radii, the proper
eccentricity with respect to the z-axis, the angular velocity, the ADM mass, the rest mass, 
the corresponding (with the same rest mass) spherical compactness, the ratio of kinetic over gravitational potential energy, 
and the moment of inertia, respectively. 
To convert to cgs units, use the fact that 
$1=1.477\ {\rm km}=4.927\ \mu {\rm s} = 1.989\times 10^{33}\ {\rm g}$.}
\label{tab:models}
\end{table*}

From the left panel of Fig. \ref{fig:tov} we notice that the bifurcation point happens at larger $\GB$ or $e$ as the compactness
increases. The triaxial sequence also shrinks the larger the compactness, which intuitively means that it is harder to construct
a triaxial neutron star of large compactness. 
For incompressible fluids \cite{Rosinska2002a} in general relativity it was found that
\be
\GB_s = \GB_s^{\rm Newt} + 0.126\left(\frac{M}{R}\right)_s\left(1+\left(\frac{M}{R}\right)_s\right)
\label{eq:ToWsec}
\ee
where $\GB_s^{\rm Newt}=0.1375$ at eccentricity $e_s^{\rm Newt}=0.8127$.
Equation (\ref{eq:ToWsec}) predicts that $\GB_s=0.15$ at $(M/R)_s=0.1$ while for $(M/R)_s=0.19$, it is
$\GB_s=0.166$, which are in broad agreement with the right panel of Fig. \ref{fig:tov}.  
Notice also that the IWM formulation sligthly
overestimates $\GB_s$ as well as $e_s$ at the bifurcation point with respect to
a full solution to the Einstein equations \cite{Uryu2016a}.

The models used in this study are 
shown in the right panel of Fig. \ref{fig:tov} as blue and magenta dots. 
They constitute the most triaxial solutions along the corresponding sequences of
constant rest mass.
In Table \ref{tab:models}, these two solutions are dubbed as C010s17 (magenta corresponds 
to compactness $0.1$) and C019s08 (blue corresponds to compactness $0.19$).

We have employed the single star module of the \cocal{} code to compute the 
quasi-equilibrium solutions of this work.
This module uses the KEH method \cite{Komatsu89,Komatsu89b} on a single 
spherical patch $(r,\GU,\GP)$ with $r\in[r_a,r_b]$,
$\GU\in[0,\pi]$, and $\GP\in[0,2\pi]$, where $r_a=0$, $r_b=O(10^6 M)$
(no compactification used), to achieve 
convergence through a Green's function iteration.  The
grid structure in the angular dimensions is equidistant but not in the
radial direction.  The definitions of the grid parameters can be seen
in Table \ref{tab:grids_param}, along with the specific values used here.

\begin{table}
\begin{tabular}{lll}
\hline
\hline
$r_a=0$         &:& Radial coordinate where the radial grids start.       \\
$r_b=10^6$      &:& Radial coordinate where the radial grids end.     \\
$r_c=1.25$      &:& Radial coordinate between $r_a$ and $r_b$ where   \\
&\phantom{:}  & the radial grid spacing changes.   \\
$N_{r}=384$     &:& Number of intervals $\Dl r_i$ in $r \in[r_a,r_{b}]$. \\
$\Nrf=128$      &:& Number of intervals $\Dl r_i$ in $r \in[r_a,1]$. \\
$\Nrm=160$      &:& Number of intervals $\Dl r_i$ in $r \in[r_a,r_{c}]$. \\
$N_{\theta}=96$ &:& Number of intervals $\Dl \theta_j$ in $\theta\in[0,\pi]$. \\
$N_{\phi}=96$   &:& Number of intervals $\Dl \phi_k$ in $\phi\in[0,2\pi]$. \\
$L=12$          &:& Order of included multipoles. \\
\hline
\hline
\end{tabular}  
\caption{Summary of grid parameters used by \cocal{} to produce the triaxial models.
Note that $\Nrf=128$ is the number of points across the largest star radius. }
\label{tab:grids_param}
\end{table}

\subsection{Evolutions}
\label{sec:evol}

For the evolution we use the \baikal~\cite{Ruchlin2017}
code, which solves the Einstein field
equations in the BSSN formalism and the \illinois{}~\cite{Etienne2015,Noble2006} to evolve fluid quantities.
The code is
built on the \cactus{} infrastructure and uses \carpet{}~\cite{Schnetter:2003rb} 
for mesh refinement, which allows us to focus numerical resolution on
the strong-gravity regions, while also placing outer boundaries at
large distances well into the wave zone for accurate GW extraction and
stable boundary conditions.  
The evolved geometric variables are the
conformal metric $\tilde{\GG}_{ij}$, the conformal factor $\GP$,
($\GG_{ij}=e^{4\GP}\tilde{\GG}_{ij}$), the conformally-rescaled,
tracefree part of the extrinsic curvature, $\tilde{A}_{ij}$, the trace
of the extrinsic curvature, $K$, and three auxiliary variables
$\tilde{\Gamma}^i=-\pd_j\tilde{\GG}^{ij}$, a total of $17$ functions.
For the kinematical variables we adopt the puncture gauge conditions
~\cite{Baker:2005vv,Campanelli06}, 
which are part of the family of gauge conditions using an advective 
``1 + log'' slicing for the lapse, and a 2\textsuperscript{nd} order
``Gamma-driver'' for the
shift ~\cite{Alcubierre2003}.
 
The equations of hydrodynamics are solved in conservation-law form adopting
high-resolution shock-capturing methods \cite{DelZanna2002}.  The
primitive, hydrodynamic matter variables are the rest mass density,
$\GR$, the pressure $p$ and the coordinate three velocity
$v^i=u^i/u^0$. The enthalpy is written as $h=1+e+p/\GR$, and
therefore the stress energy tensor is
$T_{\GA\GB}=\GR hu_\GA u_\GB+p g_{\GA\GB}$. Here, $e$ is the specific internal
energy\footnote{This should not be confused with the eccentricity Table \ref{tab:models}.}.
%

The grid structure used in these evolutions is summarized in Table
\ref{tab:evol_param}.  Typically we use five refinement levels with the
innermost level half-side length being approximately $\sim 1.5$ times
larger than the radius of the star in the initial data ($R_x$). We use
$240 \times 240 \times 240$ cells for the innermost refinement level,
which means that we have approximately 160 points across the neutron
star largest diameter. (For the initial data construction we used 256 
points across the largest neutron star diameter.) 
For the innermost refinement level this implies a $\Delta x
\sim 5.53\times10^{-3}$ (C010s17) and $\Delta x \sim 5.60\times10^{-3}$
(C019s08).  This number of points was necessary
in order to have accurate evolutions of such stiff equations of state
($\Gamma=4$) which present a challenge for any evolution code.

\begin{table*}
\begin{tabular}{cccccccccc}
\hline
\hline
Model & Grid hierarchy & $dx$ & $N$  \\  
\hline
C010s17 & $\{8.49,4.24,2.12,1.06,0.531\}$ & $8.84\times10^{-2}$ & $80$  \\
C019s08 & $\{8.59,4.30,2.15,1.07,0.537\}$ & $8.95\times10^{-2}$ & $80$  \\
\hline
\hline
\end{tabular}
\caption{Grid parameters used for the evolution of each
model. Parameter $N$ corresponds to the number of points used to
cover the largest radius of the star. Parameter $dx$ is the step interval in the coarser
level. }
\label{tab:evol_param}
\end{table*}

\begin{figure}[H]
\begin{center}
\includegraphics[width=6.5 cm]{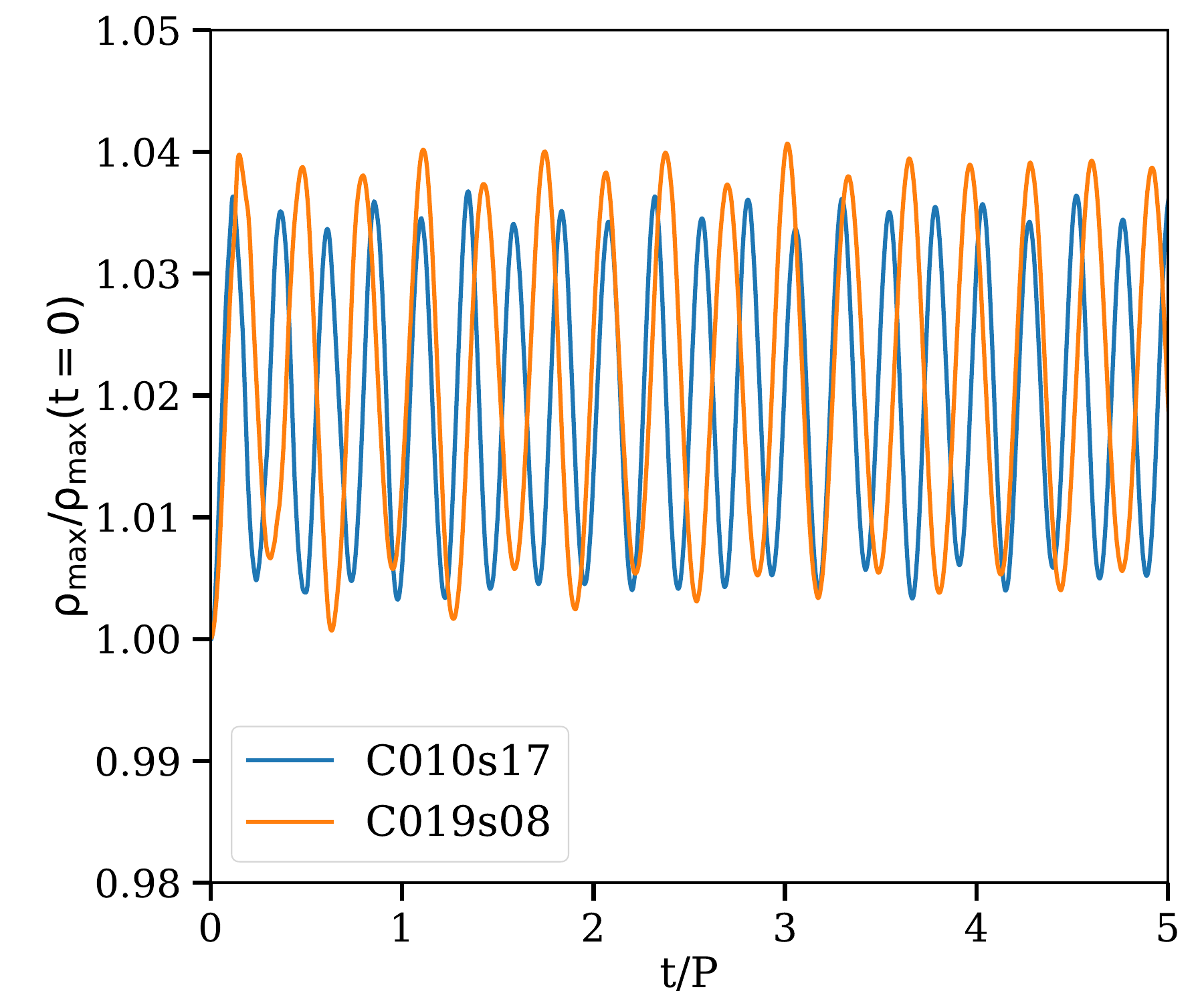}
\caption{Behavior of maximum density for models C010s17 and C019s08 under a $5\%$ pressure depletion.}
\label{fig:pres}
\end{center}
\end{figure}

\section{Results}
\label{sec:results}

We perform full general relativistic simulations of the two most triaxial models C010s17 and C019s08
under a variety of perturbations in order to probe their stability and more importantly their fate
especially with respect to their nonaxisymmetric shape. 
As a first test for dynamical stability we evolve these triaxial models by applying a $5\%$
pressure depletion in their interior. Note that the dynamical timescale is a fraction of the 
period $P$ of rotation of any system 
\be
\frac{t_d}{P} = \frac{t_d}{M}\frac{M}{P} \sim \frac{1}{\Omega M}\frac{M}{P} \sim 0.16 \ .
\label{eq:td}
\ee
In Fig. \ref{fig:pres} the maximum (central) density oscillations versus time are 
shown for five rotation periods.
Overall, both models in Table \ref{tab:models} show the
same oscillatory behavior when we pressure-deplete them, thus they
are stable against quasiradial perturbations on dynamical timescales. Since these
are the most triaxial models along a sequence of constant rest mass, the presented triaxial 
sequence (quasiequilibria along magenta or blue lines in right panel of Fig. \ref{fig:tov})
would also be stable. 

Having established the dynamical stability against quasiradial perturbations,
we now focus on the velocity flow of the triaxial figures of quasiequilibrium and 
investigate how it affects their global hydrodynamical stability. Let us recall that in Newtonian gravity
the velocity of a Riemann-S ellipsoid in the inertial frame is 
\be
v^i = \left(  \left( \frac{R_x}{R_y}\Lambda - \Omega\right)y, 
              \left(-\frac{R_y}{R_x}\Lambda + \Omega\right)x, 0 \right) \ ,
\label{eq:vriemann}
\ee
where $\Omega$ is the angular velocity of the ellipsoid and $\Lambda$ the
angular frequency of the internal fluid circulation
\cite{Chandrasekhar69c,Lai93}. When there is no internal fluid
circulation ($\Lambda=0$) the fluid velocity describes the velocity field of a
Jacobi ellipsoid with vorticity $\GZ=-2 \Omega$, while when there is no angular
velocity ($\Omega=0$) the fluid velocity describes the velocity field of a
Dedekind ellipsoid with vorticity $\GZ=-(\frac{R_x}{R_y}+\frac{R_y}{R_x})\Omega$.

\begin{figure}[H]
\begin{center}
\includegraphics[width=6.5 cm]{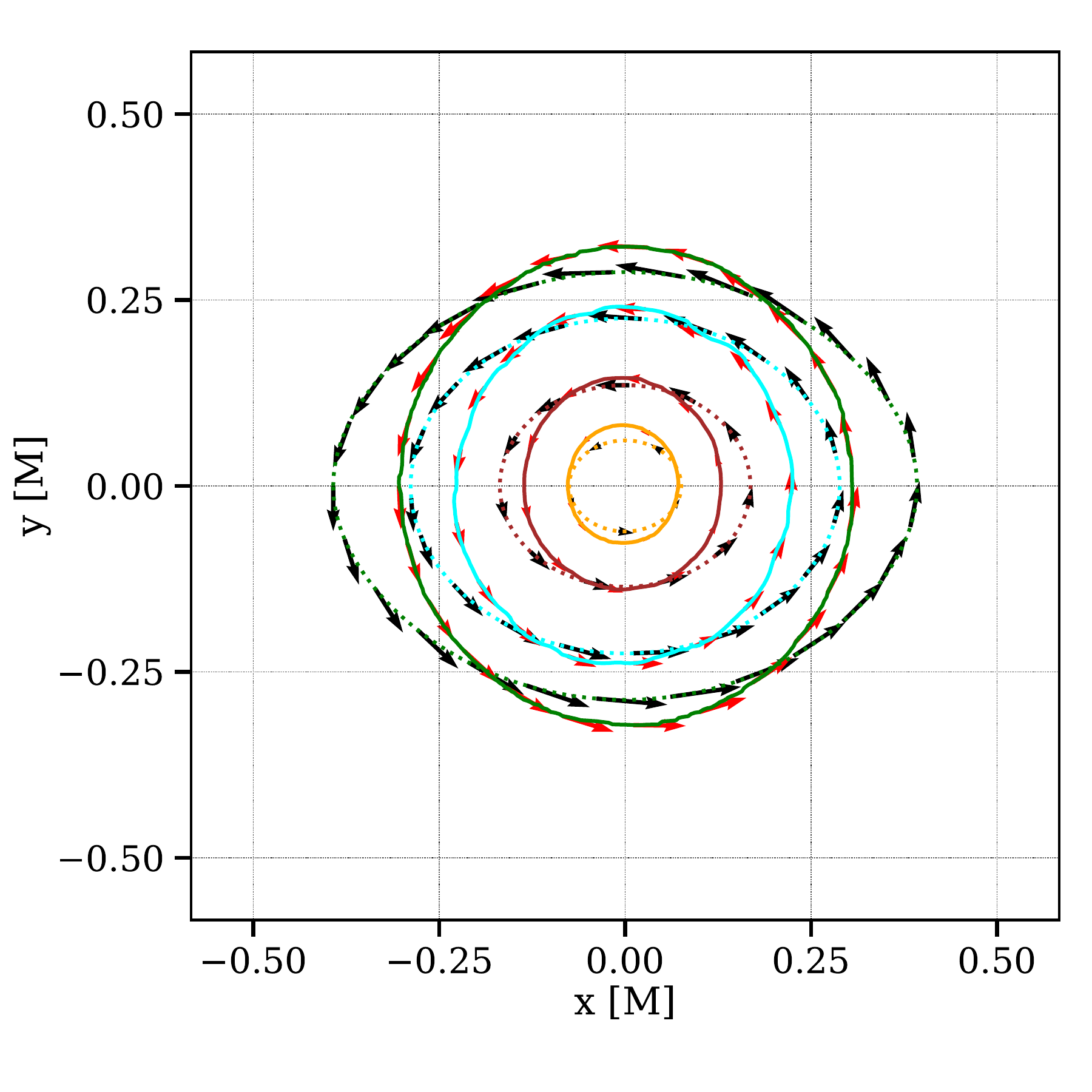}
\includegraphics[width=6.5 cm]{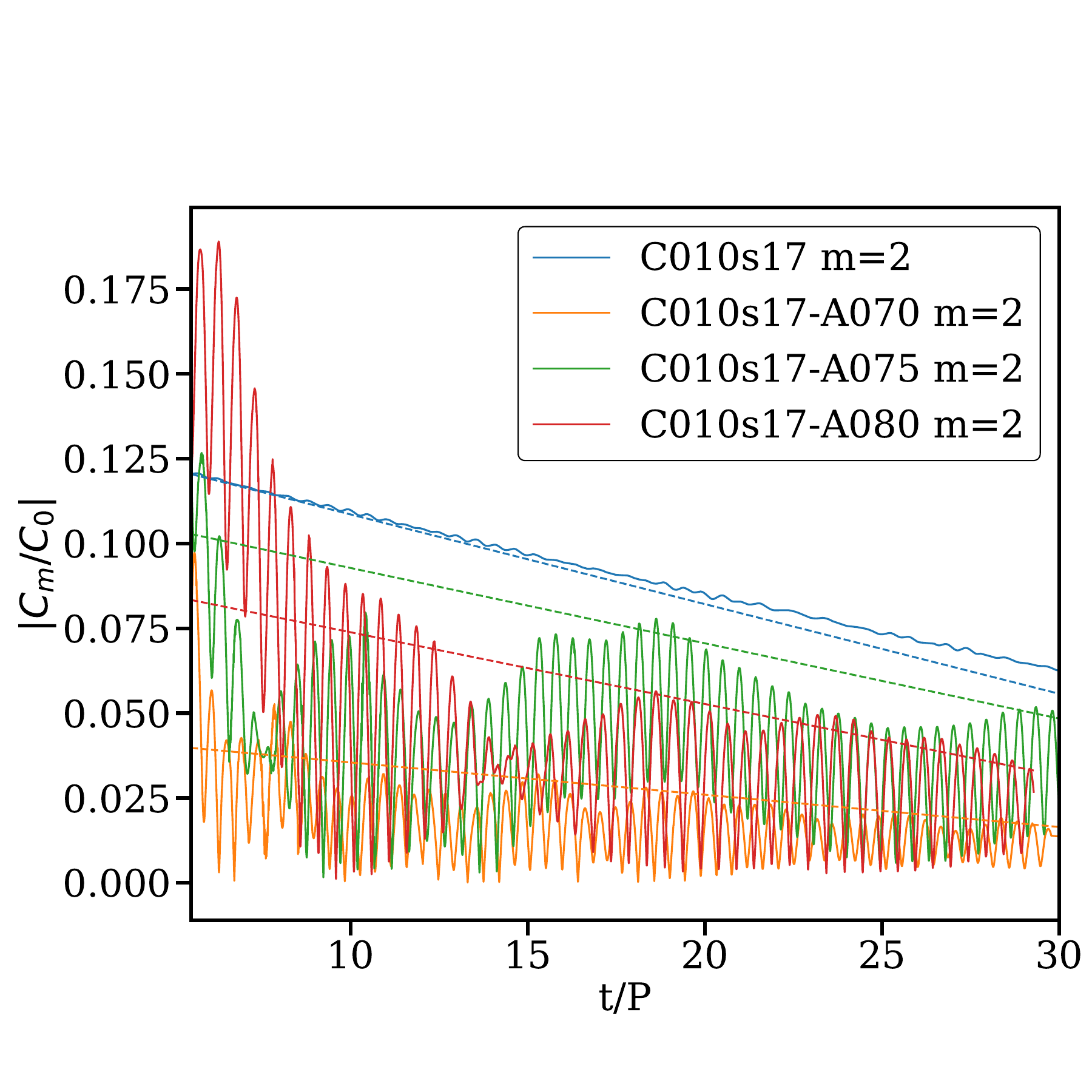}
\caption{Left panel: Solid (dotted) colored lines are density isocontours at $t/P=30.6$ ($t/P=0$) for the
model C010s17-A075. The velocity field (red arrows for $t/P=30.6$ and black ones for $t/P=0$) is also shown. 
Right panel: The $m=2$ mode amplitude for triaxial model
C010s17 as well as for all its velocity perturbed models. Dashed lines are linear fits.}
\label{fig:velocitypert}
\end{center}
\end{figure}

Since models C010s17 and C019s08 are the analogues of Jacobi ellipsoids in general relativity
with a compressible equation of state, we replace 
their velocity flow field by the corresponding one of Dedekind ellipsoids. 
By substituting in Eq. (\ref{eq:vriemann})
$\Omega=0$ and $\Lambda=\Omega$, we find that the star significantly destabilizes, 
hence we follow the procedure below. 
First we identify the contours of constant rest-mass density, and construct their tangential directions. 
We then assign at each point a velocity whose direction is the one computed from the tangent to the isocontours 
and its magnitude is given by $|\Lambda| ((y R_x/R_y)^2 + (x R_y/R_x)^2)^{1/2}$, where $\Lambda$ is a free parameter. In this way we ensure that the velocity field is consistent with the density gradients and 
only its magnitude can cause significant deformations.

Setting $A=\Lambda/\Omega$ we find that for the model C010s17 and $A=0.7,\ 0.75,\ 0.8$, the rotational motion of the triaxial figure 
is greatly reduced but significant nonaxisymmetric oscillations are present. We refer to these evolutions as 
C010s17-A070, C010s17-A075, and C010s17-A080 respectively.
In the left panel Fig. \ref{fig:velocitypert} dotted colored lines show the density 
isocontours at $t=0$ with the velocity field constructed in the way explained above 
for the C010s17-A075 model. Also shown are the isocontours and velocity field 
at the end of our simulations at $t/P=30.6$. 
In the right panel of Fig. \ref{fig:velocitypert} we plot the $m=2$
nonaxisymmetric mode amplitude for all four models (C010s17 and it perturbed models) 
\be
C_m = \int \GR u^t e^{i m\GP} \sqrt{-g} d^3x \ ,
\label{eq:Cm}
\ee
normalized by the rest mass of the system.
For the nonperturbed case, C010s17, the amplitude of $C_2$ is monotonically decreasing until the 
end of our simulations. This is due to the emission of gravitational waves and its loss of energy 
and angular momentum, which make the neutron star more axisymmetric. Notice that a nonrotating triaxial 
ellipsoid that preserved its shape (as an equilibrium Dedekind ellipsoid) 
would have a constant $C_2$ for all times in a simulation.
The perturbed cases C010s17-A070, C010s17-A075, and C010s17-A080 show an initial increased bar mode 
which decays in different ways. In all four cases we also plot with
dashed lines linear fits. As we can see all three perturbed
models evolve towards axisymmetry with different rates. The model that clearly shows the least amount 
of triaxiality {\it change} (for $t/P\gtrsim 10$) is C010s17-A070. 

\begin{figure}
\includegraphics[width=13cm]{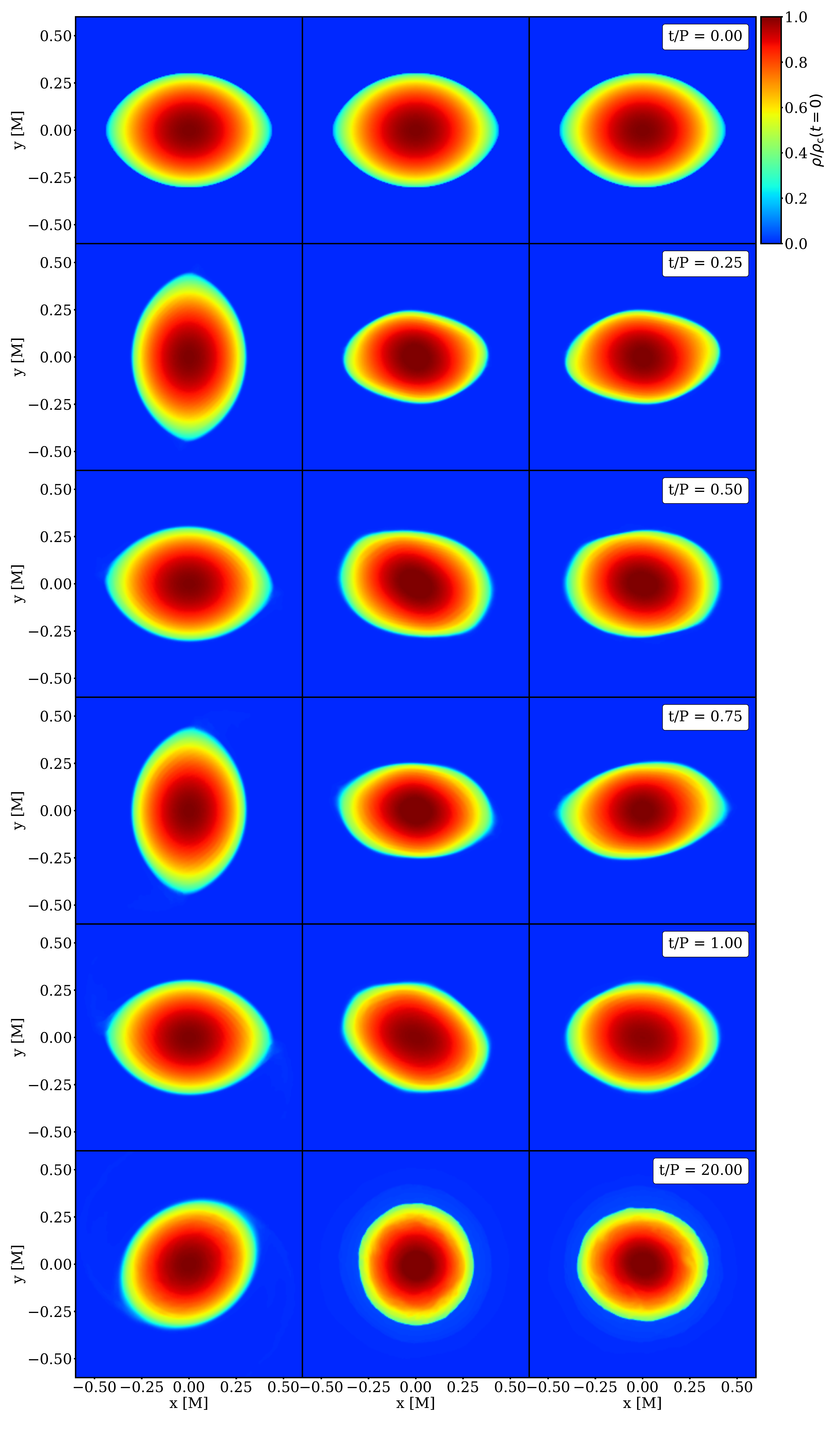}
\caption{Density plots for models C010s17 (left column), C010s17-A070 (middle column), C01s17-A075 (right column) 
at different time instances (each row).}
\label{fig:C010s17_density_panels}
\end{figure}

In order to appreciate the overall motion of these ellipsoidal figures, we plot in 
Fig. \ref{fig:C010s17_density_panels}, left column, the density profile of the nonperturbed 
case C010s17 at a select number of times $t/P=0,\ 0.25,\ 0.5,\ 0.75,\ 1.0,\ 20.0$. In the
middle column we plot for the same times model C010s17-A070, while at the right column we 
plot model C010s17-A075. As can be seen from the first five instances ($t/P\lesssim 1$) where 
the nonperturbed model C010s17 makes one complete revolution, models C010s17-A070 and 
C010s17-A075 barely rotate while they exhibit nonaxisymmetric deformations. By the end of our 
simulations at $t\sim 30 P$ all models remain nonaxisymmetric (although less than at $t=0$) and 
continue to oscillate mildly.

\begin{figure}
\begin{center}
\includegraphics[width=6.5cm]{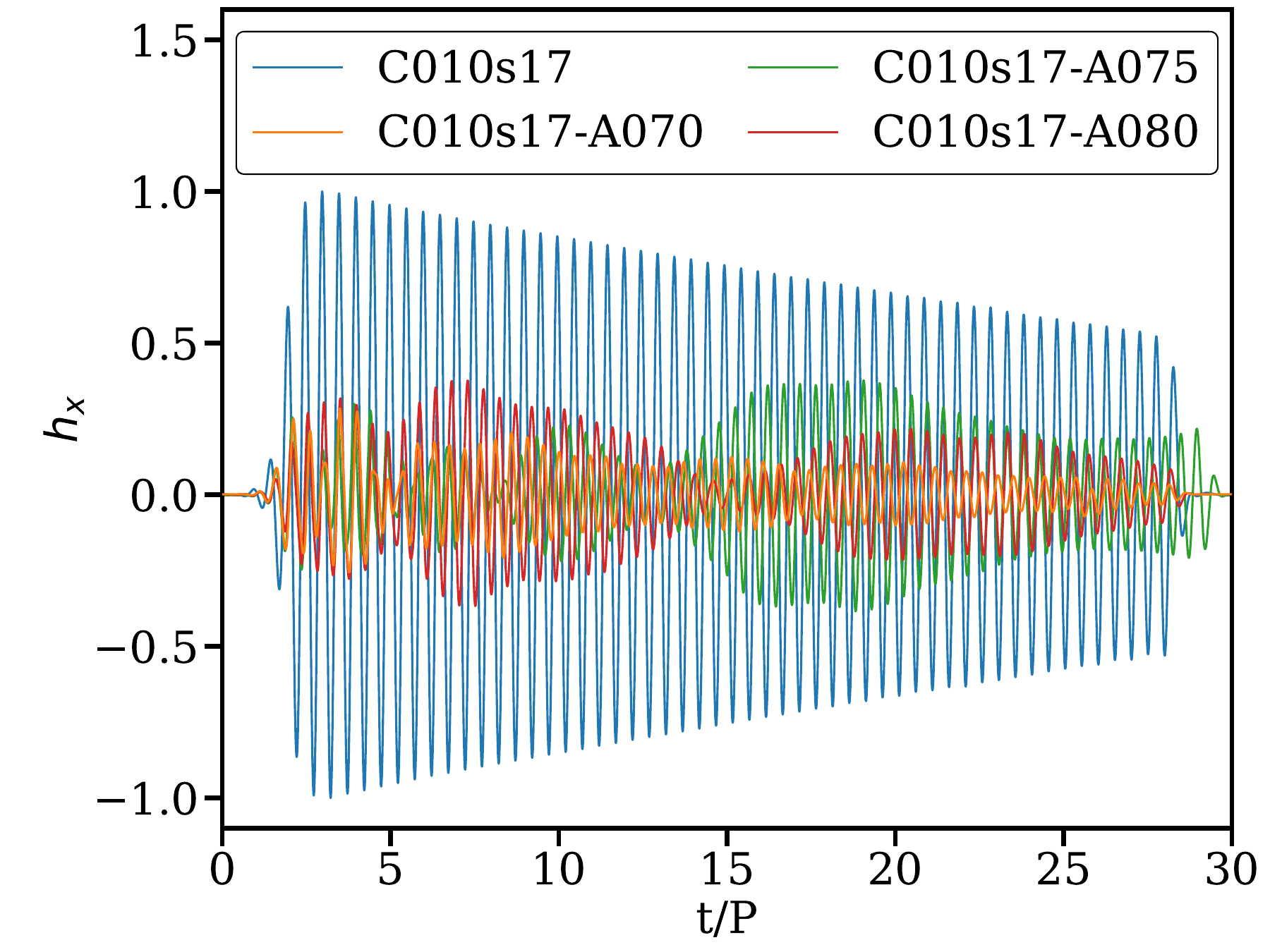}%
\includegraphics[width=6.5cm]{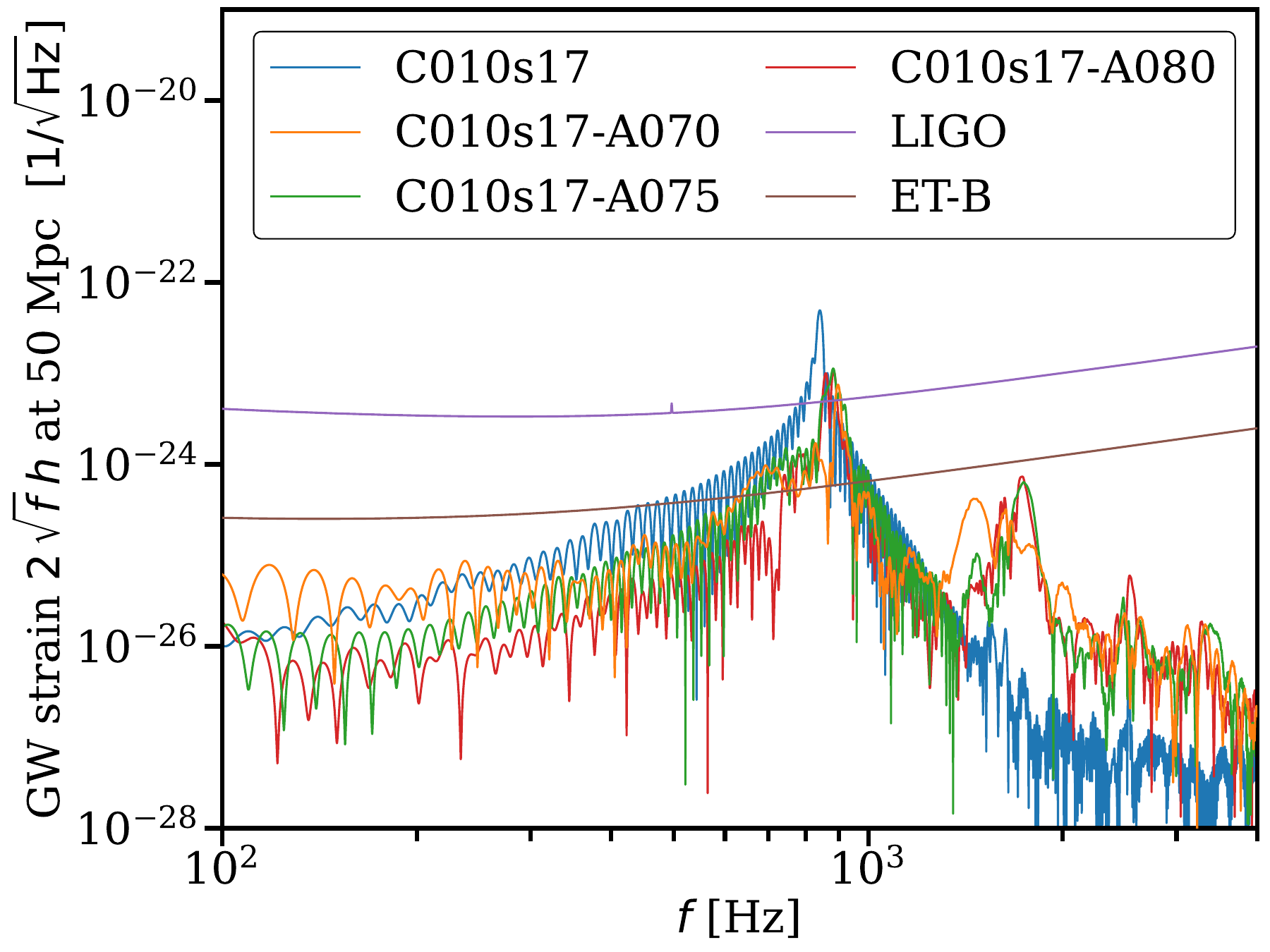}
\caption{Left panel: Gravitational wave strain normalized to the maximum value of the unperturbed model C01s17.
Right panel: Power spectrum of the C010s17 models for a $1.4\,M_\odot$ triaxial star at a distance of
50 Mpc. }
\label{fig:C01s17_hx}
\end{center}
\end{figure}

In the left panel of Fig.~\ref{fig:C01s17_hx} we plot 
the gravitational wave strain $h_\times$ for the C010s17 models
following \cite{Reisswig:2011}. It is obvious that the Dedekind-like velocity flow decreased the 
gravitational wave signature of the triaxial figures by less than half, even at early times. 
Consistent with Fig. \ref{fig:C010s17_density_panels} and the left panel of Fig. \ref{fig:velocitypert}
we see that the model with the least amount 
of gravitational wave content is C010s17-A070 (orange line) whose strain exhibits a decrease 
at $\sim 20\%$ of the original C010s17 case (blue line). Note that since the perturbed models are 
barely rotating these high frequency gravitational waves are due to the hydrodynamical flow 
perturbations in the neutron stars, which showcases the importance of accurate hydrodynamical
modeling for the understanding of a physical system through gravitational waves.
In the right panel  of Fig.~\ref{fig:C01s17_hx} we plot the power spectrum of the C010s17 models
scaled for a $1.4\, M_\odot$ triaxial neutron star mass at $50\, \rm Mpc$, 
along with the noise curves for Advanced LIGO's design
sensitivity~\cite{ligonoise:web} and the ET-B configuration of Einstein Telescope~\cite{etnoise:web}.
The peak frequency of the unperturbed C010s17 model (blue curve) at $\sim 840\, \rm Hz$ is
consistent with its orbital angular frequency ($\Omega/\pi$) and in principle detectable by
Advanced LIGO. Consistent with the left panel, the power spectrum of the perturbed models is weaker 
but still detectable from next generation observatories such as the Einstein Telescope.
Once again, the similarities between the different curves show that the star's rotation can be degenerate 
with the fluid flow in the frequency domain.

The computational experiment performed with model C010s17 was repeated for the more compact 
model C019s08. We found that for almost any value of the parameter $\Lambda$ that we used the
star was highly destabilized. For a select set of values (e.g. $A=0.3,\ 0.4$) where the star
survived, its rotation rate was unaffected and the its gravitational wave content was not
reduced (actually the gravitational waves became more complicated due to the induced
oscillations). Thus we were unable to create Dedekind-type of flows for these
highly relativistic and compressible objects. One way to probably improve our treatment
is to use the relativistic magnitude of the velocity $\sqrt{\GG_{ij}v^iv^j}$ instead
of the Newtonian one used here. We plan to examine this problem in the future.

\section{Conclusions}

We constructed constant rest-mass sequences of triaxial uniformly rotating neutron stars
with a compressible equation of state in general relativity. 
We examined the stability of the most triaxial members of these sequences founding them
stable against radial and nonaxisymmetric perturbations. These quasiequilibria are the
analogs of the incompressible Jacobi ellipsoids in Newtonian gravity. Jacobi ellipsoids 
are congruent to their Dedekind counterparts with no internal motion. A Jacobi ellipsoid
with angular velocity $\Omega$ has the same principal axes as the Dedekind ellipsoid
with vorticity $\GZ=(\frac{R_x}{R_y}+\frac{R_y}{R_x})\Omega$.
In general relativity this picture may be different. Here we simulated a Dedekind-type 
of flow in an Jacobi-type relativistic figure of quasiequilibrium. We found that for
small compactness ($0.1$) the triaxial neutron star evolves to a Riemann-S-type of
ellipsoid with minimal rotation and gravitational wave emission. On the other hand,
our high compactness ($0.19$) triaxial model although similarly dynamically stable, 
was unstable to almost all Dedekind-type of flows that we tried. An important
product of this investigation is the influence of a hydrodynamical fluid flow on the
generation of gravitational waves and therefore the parameter estimation of a certain
physical system.

\section{Acknowledgements}
The simulation and analysis tasks for this manuscript were performed at NCAR-Wyoming Supercomputing Center (NWSC) via WRAP allocation WYOM0144 ``Numerical simulations of rotating neutron stars with Einstein Toolkit''.
This work used Stampede2 at TACC through allocation PHY160053 from the Extreme Science and Engineering Discovery Environment (XSEDE), which was supported by National Science Foundation grant number \#1548562~\cite{towns:2014}.
This work used Stampede2 at TACC through allocation PHY160053 from the Advanced Cyberinfrastructure Coordination Ecosystem: Services \& Support (ACCESS) program, which is supported by National Science Foundation grants \#2138259, \#2138286, \#2138307, \#2137603, and \#2138296~\cite{boerner:2023}.
Plots were produced using matplotlib~\cite{Hunter:2007aa,Thomas:2021mpl} and kuibit~\cite{Bozzola:2021hus}.

\section*{Funding}
A. Tsokaros is supported by National Science Foundation Grants 
No. PHY-2308242 and No. OAC-2310548 to the University of Illinois at Urbana-Champaign. 
R. Haas is supported by National Science Foundation Grants No. : OAC-2103680,  
No. OAC-2004879, No. OAC-2005572 and No. OAC-2310548  to the University of Illinois at 
Urbana-Champaign. K. Ury\=u is supported by JSPS Grant-in-Aid for Scientific 
Research (C) 22K03636 to the University of the Ryukyus.
Y. Luo is supported by 
National Science Foundation Grant No. OAC-2004879 to the University of Illinois 
at Urbana-Champaign, and the U.S. Department of Energy, Office of Science, 
Office of High Energy Physics, under Award Number DE-SC0019022 to the University of Wyoming.
Y. Luo is also supported by the Graduate Computing Scholars award from the School of 
Computing, University of Wyoming.

\section*{Conflict of interest}
 The authors declare that they have no conflict of interest.

\section*{Declarations}
The authors declare that there are no data associated with this manuscript.

\bibliographystyle{spphys}       
\bibliography{atreferences,einsteintoolkit}

\end{document}

%% file: kigou.tex
\newcommand{\beq}{\begin{equation}} 
\newcommand{\eeq}{\end{equation}} 
\newcommand{\beqn}{\begin{eqnarray}} 
\newcommand{\eeqn}{\end{eqnarray}}

\newcommand{\zD}{{\raise1.0ex\hbox{${}^{\ \circ}$}}\!\!\!\!\!D}
\newcommand{\alone}{{\raise0.5ex\hbox{${}^{\ 1}$}}\!\!\!\!\alpha}

\newcommand{\Dl}{\Delta}

\newcommand{\nalam}{\mathrel{\raise0.9ex\hbox{$^\lambda$}\mkern-14mu
\lower0.0ex\hbox{$\nabla$}}}

\newcommand{\Nrf}{{N_r^{\rm f}}}
\newcommand{\Nrm}{{N_r^{\rm m}}}

\newcommand{\zeroD}{{\raise1.0ex\hbox{${}^{\ \circ}$}}\!\!\!\!\!D}

\newcommand{\zLap}{{\raise1.0ex\hbox{${}^{\ \circ}$}}\!\!\!\!\Delta}
\newcommand{\zna}{{\raise1.0ex\hbox{${}^{\ \circ}$}}\!\!\!\!\!\nabla}
\newcommand{\zS}{{\raise1.0ex\hbox{${}^{\ \circ}$}}\!\!\!\!\!S}

\newcommand{\baikal}{\textsc{baikal}}
\newcommand{\carpet}{\textsc{carpet}}
\newcommand{\cocal}{\textsc{cocal}}

\newcommand{\cactus}{\textsc{cactus}}

\newcommand{\etoolkit}{\textsc{Einstein Toolkit}}

\newcommand{\illinois}{\textsc{Illinois GRMHD}}